\documentclass[reprint, aip, rsi, groupedaddress]{revtex4-1}
\usepackage{graphicx}
\usepackage[OT4]{fontenc}

\draft % marks overfull lines with a black rule on the right

\begin{document}

\title{Note: High Density Pulsed Molecular Beam for Cold Ion Chemistry}

\author{M.G. Kokish, V. Rajagopal}

\author{J.P. Marler}
\altaffiliation{Current address: Clemson University, Clemson, South Carolina.}
\author{B.C. Odom}
\altaffiliation{To whom correspondence should be addressed; E-mail: b-odom@northwestern.edu}
\affiliation{Department of Physics and Astronomy, Northwestern University, Evanston, IL, USA.}

\date{\today}

\begin{abstract}
A recent expansion of cold and ultracold molecule applications has led to renewed focus on molecular species preparation under ultrahigh vacuum conditions. Meanwhile, molecular beams have been used to study gas phase chemical reactions for decades. In this manuscript, we describe an apparatus that uses pulsed molecular beam technology to achieve high local gas densities, leading to faster reaction rates with cold trapped ions. We characterize the beam's spatial profile using the trapped ions themselves. This apparatus could be used for preparation of molecular species by reactions requiring excitation of trapped ion precursors to states with short lifetimes or for obtaining a high reaction rate with minimal increase of background chamber pressure.
\end{abstract}

\maketitle

In the past decade, researchers have steadily discovered an increasing number of exciting uses for trapped molecular ions. Topics ranging from cold chemistry\cite{Willitsch2008, Rellergert2011} to electron electric dipole moment searches\cite{Loh2013} using trapped molecular ions have become active areas of research, and interest in molecular ion preparation has led to a number of reaction rate measurements.\cite{DePalatis2013,Roth2008,Chang2013,Drewsen2000} Cold diatomic ions are typically formed by preparing laser-cooled alkaline earth metal atomic ions under ultrahigh vacuum (UHV) conditions and reacting them with neutral gas leaked into the chamber. Product formation rates are often limited by the amount of neutral gas that can be introduced, which is constrained by pumping speed and trap de-loading. As an alternative neutral precursor source, a molecular beam provides a key relative advantage: a beam can be used to concentrate the flux of neutral reactants into the ion trap, increasing product formation rates over short time scales.\cite{Deckers1963,Collings2009} As a result, the total amount of neutral gas required to achieve the desired number of molecular ions is reduced. Here we demonstrate this concept in the context of cold trapped ions by using a pulsed molecular beam source to form target molecular ion species from atomic ion precursors, and we use the trapped ions to directly measure the density profile.

To characterize our beam we exploit the previously-studied Ba$^{+}$ + N$_{2}$O $\rightarrow$ BaO$^{+}$ + N$_{2}$ reaction.\cite{Roth2008} Using a linear Paul trap (2z$_{0}$ = 17.8 mm, r$_{0}$ = 4.6 mm), we prepare a Doppler-cooled Coulomb crystal\cite{Drewsen1998} of Ba$^{+}$ ions in the path of the molecular beam. The trap is housed in an ultrahigh vacuum chamber (Kimball Physics 6.0$^{''}$ spherical octagon) held at approximately 5$\times$10$^{-11}$ torr. We form Ba$^{+}$ in the trap by selectively photoionizing the $^{138}$Ba isotope from a barium oven situated below the trap.\cite{Steele2007} The trap is operated with a 3 MHz radiofrequency (rf) drive; typical rf and end cap voltages are between 1.0-1.2 kV and 5-15 V, respectively. The ions are Doppler cooled with overlapping cooling (493 nm) and repumping (650 nm) lasers along the trap axis and are imaged with an EMCCD camera.

\begin{figure}
\includegraphics[scale=0.45]{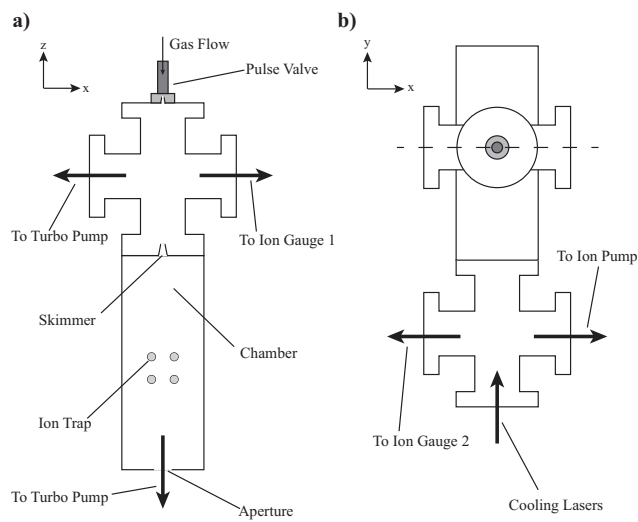}
\caption{\label{fig:app}a) Reactant gas enters through the pulse valve, is collimated by the skimmer, passes through the ion trap and exits through an aperture opening to the turbo pump. b) Top view of the apparatus with the cross section (a) taken along the dashed line.}
\end{figure}

\begin{figure}
\includegraphics[scale=0.31]{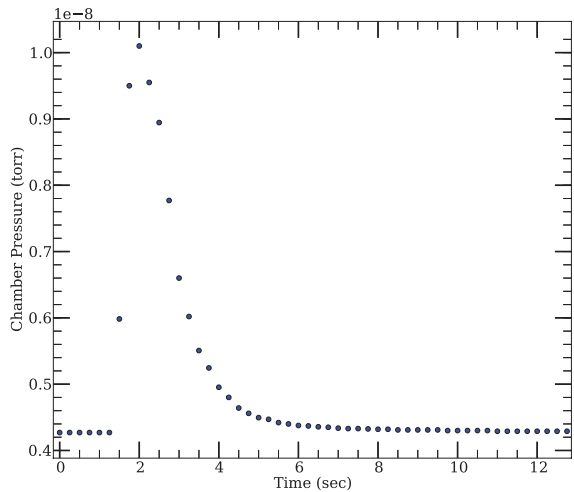}
\caption{\label{fig:pres}Temporal pressure profile after driving the pulse valve for 200 $\mu$s seen here beginning at 1.25 s.}
\end{figure}

The N$_{2}$O reactant gas is introduced into a differentially pumped configuration (Fig.~\ref{fig:app}) through a previously-characterized solenoidal pulse valve (Parker series 9, 0.5 mm, conical nozzle)\cite{Brandi2011} with typical backing pressures between 6-10 psig. We drive the valve opening with a 200 V pulse for 210 $\mu$s. Upon opening the poppet, the gas supersonically expands into a differential pumping chamber and is collimated using a 1 mm skimmer located 137 mm downstream, which provides passage into the main reaction chamber (where the trap is located). The distance between the pulse valve opening and the trap center is approximately 225 mm. The differential pumping chamber is connected to a turbo pump, while the reaction chamber is pumped with an ion pump and a turbo pump with a small aperture. The reaction chamber exit leading to the turbo pump is fixed opposite the skimmer entrance with an aperture of 2 mm, effectively creating a reverse differential pumping configuration. Figure \ref{fig:pres} shows the temporal profile of the reaction chamber pressure after a typical pulse of gas, measured from Ion Gauge 2 (Fig.~\ref{fig:app}).
\begin{figure}
\includegraphics[scale=0.30]{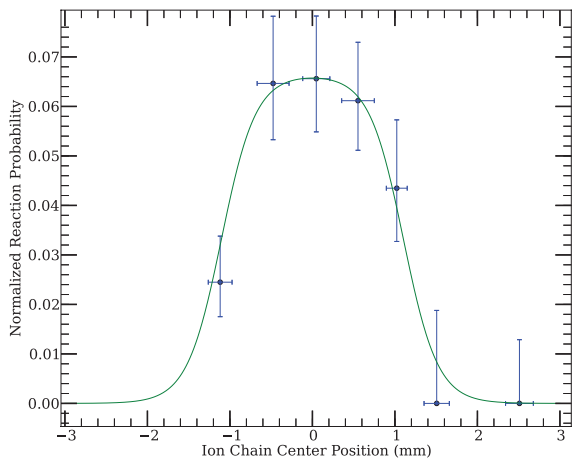}
\caption{\label{fig:prof}Reaction probability measurements with the ion crystal at different positions along the trap axis. The x uncertainty is given by the average ion crystal length for each bin. The y uncertainty refers to the 1$\sigma$ confidence intervals for a Poisson process.\cite{Note2}}
\end{figure}

\begin{figure}
\includegraphics[scale=0.9]{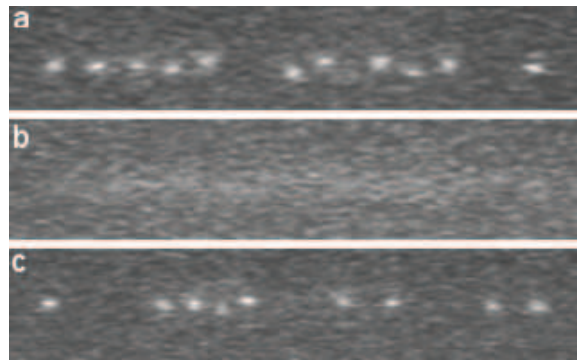}
\caption{\label{fig:vis} Typical measurement with ions in the path of the beam. a) Initial count of 11 ions. b) Melted ion string immediately after pulse of neutral gas. c) Final count of 9 bright ions.}
\end{figure}

Ba$^{+}$ ions fluoresce (bright) during Doppler cooling while the sympathetically cooled BaO$^{+}$ ions do not (dark). In order to characterize the reaction rate, we measure the number of bright ions before and after a pulse of N$_{2}$O. The ratio of the number of new dark ions to the original number of bright ions represents the reaction probability per ion per pulse. Though highly exothermic, the reaction products remain trapped because the ion trap potential well is deeper than the reaction enthalpy. Using this ratio, we quantitatively characterize the molecular beam profile in the trap's axial direction by changing the position of the target ions and measuring the reaction probability after a pulse of neutral gas. A typical string of 5-40 ions each separated from one another by approximately 15 $\mu$m is moved by varying the ion trap's axial offset voltage. Figure \ref{fig:prof} shows the measured beam profile. The FWHM beam width (2.2 mm) and edge width (0.2 mm) were estimated by fitting to a bi-exponential as in Ref~\onlinecite{Guevremont2000}. Assuming a conical expansion with the valve's conical angle of 45$^{\circ}$, a beam width of 1.5 mm is expected at the trap center, in fairly good agreement with our measurement. Imperfections in the skimmer construction can change its behavior to more of an effusive source, which would broaden the beam.\cite{pauly2000atom} The beam-like character of the gas jet is also qualitatively established by fluorescence transient dynamics. When the ions are in the path of the beam, the ion crystal melts, a result of heating induced by collisions with the neutral gas. During this time, the overall drop in fluorescence as well as ion de-localization is observed with the EMCCD (Fig.~\ref{fig:vis}). This behavior is not observed for ion crystals positioned a few millimeters from the beam center, confirming that high neutral gas pressures are truly localized.

In a single pulse, the number of ions reacted $\Delta N$ is approximately given by:
\begin{eqnarray}
\Delta N = k'N_0 \tau
\label{eq:one}
\end{eqnarray}
\begin{eqnarray}
k' = k\rho
\label{eq:two}
\end{eqnarray}
where $N_{0}$ is the starting number of fluorescing atomic ions and $\tau$ is the pulse duration. The neutral gas density is much larger than that of the trapped ions and is unaffected by the reactions. Eq.~\ref{eq:two} defines the pseudo-first order rate constant $k^{'}$, where $\rho$ is the number density of the neutral gas and $k$ is the bimolecular rate constant. Energetically, Ba$^{+}$ must be in the excited state in order to react.\cite{Roth2008} Its average excited state population is assumed to remain constant over the timescale of the pulse, because Ba$^{+}$ is constantly excited by the always-on Doppler cooling lasers. Using our measured peak reaction probability ($\Delta N / N_0$) of 0.066 per ion per pulse and the previously measured rate constant (0.016 s$^{-1}$),\cite{Roth2008} we estimate a number density ($\rho$) of 10$^{11}$/cm$^{3}$ over a time period of 0.5-1 ms.\cite{Note1} For comparison to a typical leak valve scenario, our reaction rate corresponds to leaking in neutral gas at 10$^{-9}$ torr ($\sim$10$^{7}$/cm$^{3}$) for approximately 2 seconds. As an illustrative example, one can consider reaction rate measurements outside the beam as being equivalent to those from a leak valve; the ratio of the peak reaction rate to that outside the beam represents the overall reduction in gas. For our apparatus we observe no reactions outside the beam. Using Poisson statistics, this allows us to set a 68\% confidence-limit lower bound of five on this ratio.

In order to achieve higher reaction rates, higher pulse valve backing pressures could be managed with an improved reverse differential pumping configuration. A majority of the high-density molecular beam that passes through the trap flows directly into the exit on the opposite side, with the aperture mitigating backscattered gas re-entering the reaction chamber. By using an aperture diameter a few times larger than in the current design or by achieving better beam collimation, we could reduce the transient background pressure from each pulse (Figure \ref{fig:pres}.)  Under these improved conditions, we could then operate our pulse-valve backing pressure up to 1250 psi and increase the reaction probability per pulse.  We note that operating at higher backing pressures leads to increased beam-broadening,\cite{pauly2000atom} which would need to be compensated by aperture size or beam collimation technique.

In summary, we have assembled a pulsed molecular beam in conjunction with laser-cooled trapped ions. In contrast to a previous pulsed beam experiment,\cite{Chang2013} we manipulate the positions of the ions to directly characterize the beam's spatial density profile. We show that higher reaction rates can be achieved while avoiding high background chamber pressures. These results serve as a proof of principle demonstration for broadening the scope of molecular ion species for which cold chemistry can be investigated. Pulsed molecular beams are particularly advantageous when low duty cycles can be tolerated,\cite{Cross1982} such as synchronizing gas pulses with trapped molecular ions having short excited state lifetimes. Observing such reactions would often be unfeasible using a leak valve due to the unacceptably high chamber pressures required.

This work was supported by AFOSR grant no. FA9550-13-1-0116 and NSF grant no.
PHY-1309701.

\bibliography{beam}

\end{document}